\documentclass[fleqn,10pt]{wlscirep}

\title{Tunable plasmonic resonances in highly porous nano-bamboo Si-Au superlattice-type thin films}
\author[1,*]{Ufuk K{\i}l{\i}\c{c}}
\author[1,3]{Alyssa Mock}
\author[1]{Ren\'e Feder}
\author[1]{Derek Sekora}
\author[1]{Matthew Hilfiker}
\author[1]{Rafal Korlacki}
\author[1]{Eva Schubert}
\author[1]{Christos Argyropoulos}
\author[1,3,4]{Mathias Schubert}

\affil[1]{Department of Electrical and Computer Engineering, University of Nebraska-Lincoln, Lincoln, NE 68588, USA}
\affil[2]{Leibniz Institute for Polymer Research, Dresden, Germany}
\affil[3]{Department of Physics, Chemistry, and Biology, Link{\"o}ping University, 58183 Link{\"o}ping, Sweden}
\affil[]{ufuk.kilic@huskers.unl.edu, christos.argyropoulos@unl.edu, URL: http://ellipsometry.unl.edu}

\begin{abstract}
We report on fabrication of spatially-coherent columnar plasmonic nanostructure superlattice-type thin films with high porosity and strong optical anisotropy using glancing angle deposition. Subsequent and repeated depositions of silicon and gold lead to nanometer-dimension subcolumns with controlled lengths. The superlattice-type columns resemble bamboo structures where smaller column sections of gold form junctions sandwiched between larger silicon column sections (``nano-bamboo''). We perform generalized spectroscopic ellipsometry measurements and finite element method computations to elucidate the strongly anisotropic optical properties of the highly-porous nano-bamboo structures. The occurrence of a strongly localized plasmonic mode with displacement pattern reminiscent of a dark quadrupole mode is observed in the vicinity of the gold subcolumns. We demonstrate tuning of this quadrupole-like mode frequency within the near-infrared spectral range by varying the geometry of the nano-bamboo structure. In addition, coupled-plasmon-like and inter-band transition-like modes occur in the visible and ultra-violet spectral regions, respectively. We elucidate an example for the potential use of the nano-bamboo structures as a highly porous plasmonic sensor with optical read out sensitivity to few parts-per-million solvent levels in water.

\end{abstract}
\begin{document}

\flushbottom
\maketitle

\thispagestyle{empty}

\section*{Introduction}

\begin{figure}[hbt]
\centering
\includegraphics[width=.9\textwidth]{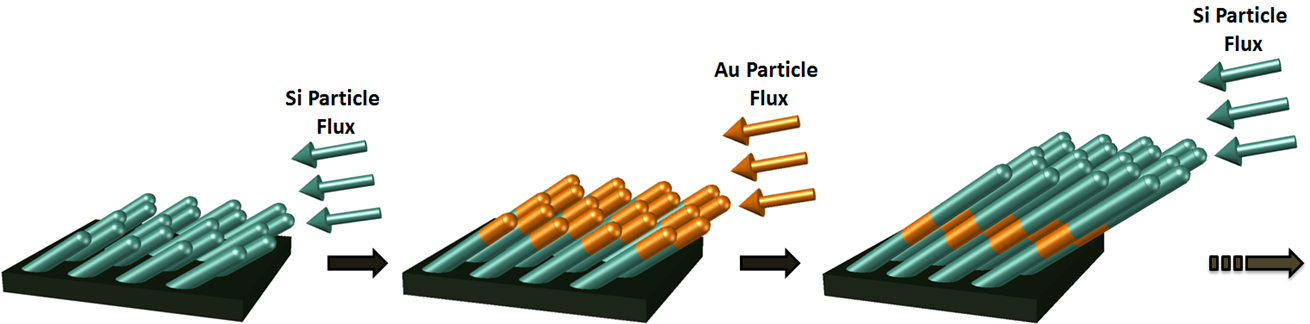}
  \caption{Schematic representation of the fabrication of slanted columnar heterostructure thin films (SCHTF), reminiscent of bamboo-like nanostructures (``nano-bamboo''), by subsequent electron-beam glancing angle deposition (GLAD) fabrication of columnar nanostructures from Si and Au sources.}
\label{fabricationSCHTF}
\end{figure}

Unraveling the mechanisms that control the optical properties of highly porous, periodic, three-dimensional (3D) arrangements of nanoplasmonic structures can offer new avenues for the development of nanoplasmonic sensors with high penetrability for liquid and gaseous substances. Periodic nanoplasmonic structures offer extensive control of light–matter interactions and are widely exploited, for example, in optoelectronic\cite{banari1995pbte,zhang2016plamon} and photovoltaic devices,\cite{brongersma2014light} ultra-fast optical switching technologies, \cite{ren2011nanostructured, dani2009subpicosecond, ahmadivand2017optical} magnetic recording, \cite{dor2013chiral, boulle2017corrigendum} microscopy and spectroscopy, \cite{li2017plasmon, chirumamilla2017hot, neubrech2017surface,khner2017nanoantenna, xi2007optical,kneipp2010novel} surface-enhanced micro-Raman scattering,\cite{liu2018high } sensor applications,\cite{yi2016dipole} biosensing,\cite{kabashin2009plasmonic,tripp2008novel} and biophotonics.\cite{narasimhan2018multifunctional} Noble metal nanoparticles combined with dielectric materials reveal light-induced excitations known as plasmon resonances.\cite{KreibigBook} These resonances provide extremely large, highly localized electric field enhancements in the immediate vicinity of the metal nanoparticles. The plasmon frequency can be controlled by the size and shape of the metal nanoparticles.\cite{LalNatPhot2007,MillstoneNanoLett2008,RingeJPCL2012} Recently, a universal dipolar plasmon energy dependence on the metal nanoparticle size regardless of shape was reported, where the plasmon energy follows approximately inversely proportional to the longest extension of the nanoparticles.\cite{RingeJPCL2012} 

\begin{equation}\label{eq:universalplasmonlength}
E_p \approx \alpha_0 - \alpha_1 L_p,
\end{equation}

\noindent where $E_p$ is the plasmon energy in units of eV, $L_p$ is the universal plasmon length in units of nm, and $\alpha_{0}= 2.376$~eV and $\alpha_1=0.00308 $~eV/nm are constants obtained from experiment. Accordingly, when desired plasmon energies are to fall within the near-infrared region, for example, nanoparticles with few hundred nanometer extension are required.\cite{RingeJPCL2012} While the size thereby provides convenient tunability of plasmon energy, large particle sizes also cause strong optical absorption and radiation loss. Hence, near-infrared plasmonic nanoparticle architectures typically are associated with large losses, and transmission configurations are not suitable. Alternative approaches seek ordered arrangements of nanoparticles thereby creating large fractions of unoccupied volume to reduce light absorption, for example, on surfaces or in thin film form. Different fabrication techniques have been used to create highly uniform periodic three dimensional structures, such as electron beam (e-beam) lithography with phase-shift mask technology,\cite{shao2006direct} nano-imprint lithography,\cite{kim2007fabrication} nanoscale selective area epitaxy,\cite{chen2007growth} laser direct writing technique,\cite{kim2014direct,huang2017realization} and glancing angle deposition (GLAD).\cite{HawkeyeJVSTA25_2007,dick2003controlled,Schmidt_TSF2013} GLAD is a physical vapor deposition technique, which utilizes particle flux at oblique angles resulting in bottom-up growth of columnar structures with nanometer dimensions due to competitive nucleation processes, geometrical shadowing, and adatom surface diffusion limitations.\cite{HawkeyeJVSTA25_2007}  For example, ordered 2D arrangements of plasmonic wires are reported for optical read-out gas sensing applications.\cite{tittl2011palladium} Ordered 3D arrangements of dense, hexagonal closed-packed and upright positioned large (hundreds of nanometer) size Au rods fabricated using e-beam lithography and e-beam evaporation methods were investigated for their plasmonic properties.\cite{huang2016tunable} It was shown that the electric field magnitude between the Au rods can be strongly enhanced, and the plasmon frequency can be tuned by height and array period.\cite{huang2016tunable} Ordered 3D arrangements of highly porous, rectangular arranged, large (hundreds of nanometers) SiC pillars etched into a semi-insulating silicon carbide substrate using e-beam lithography revealed the aspect-ratio driven evolution of higher order, multipolar, and highly localized surface phonon polariton resonances in the infrared spectral region.\cite{ ellis2016aspect}

Slanted columnar thin films (SCTF) are well known for exhibiting strongly anisotropic optical properties over a wide spectral region due to their high degree of order.\cite{Hodgkinson_1998,Schmidt2009, HofmannAPL99_2011,Schmidt_2013,Schmidt_TSF2013} Three major effective dielectric function spectra can fully describe the anisotropic optical properties of SCTFs. To this end, the spectral behavior of the formed major polarizability functions fully depends on the choices of both the chosen material to build the columnar nanostructures and the geometry of the columns (thickness, slanting angle, inter-columnar space). Plasmonic optical properties were studied in GLAD deposited indium tin oxide SCTFs decorated with Au nanoparticles using an electroplating process.\cite{BarrancoACSAMI2015} It was observed that the Au nanoparticles increased the anisotropy of the SCTFs.\cite{ranjan2015localized,BarrancoACSAMI2015} Composites of multiple materials can be achieved, for example, by fabrication of core-shell SCTFs using atomic layer deposition,\cite{SchmidtPassivation2012} chemical vapor deposition,\cite{Wilson2015} or post-growth oxidation.\cite{Mock_2016} Alternatively, in a rather new approach, slanted columnar heterostructure thin films (SCHTFs) can be fabricated by subsequent change of the source material during the GLAD growth (Fig.~\ref{fabricationSCHTF}).\cite{Sekora_2016} 

In this paper, we discuss nanoscale SCHTFs that combine a noble metal (Au) with a dielectric material (Si) in order to create and control ordered 3D arrangements of nanoplasmonic structures.  We demonstrate that the universal plasmon length rule in Eq.~\ref{eq:universalplasmonlength} is no longer valid when plasmonic nanoparticles are arranged within a lattice structure and coherently in 3-dimensional geometries. We obtain by GLAD highly-ordered, highly-porous Au nanocolumns suspended by Si nanocolumns and study their optical properties. Figure~\ref{fabricationSCHTF} depicts the sequence of our Si-Au SCHTF fabrication, resulting in nano-bamboo structures. First, a Si SCTF is deposited. Then, the source material is switched to Au, and a Au SCTF is deposited onto the first Si SCTF. Key is the continuation of the columnar growth mode. This sequence is repeated four times and a 4 $\times$ Au-Si SCHTF is fabricated. The details about the fabrication process are given in the methods section. We perform generalized spectroscopic ellipsometry (GSE) investigations from the near-infrared to the ultra-violet spectral regions. The details about the ellipsometry approach are given in the methods section. From analysis of our ellipsometry data, we obtain the anisotropic optical functions of the SCHTFs. Of particular interest is the extinction coefficient for electric field polarization parallel to the nano-bamboo axis, where strong absorption peaks occur upon the incorporation of Au within Si-SCTFs. We observe and discuss these absorption peaks for a large set of samples with different total column lengths but at constant total Si-Au volume ratio within the nano-bamboo structures. We perform finite element method (FEM) computational modeling of the optical properties and we show the electric field distributions within the nano-bamboo structures. High-resolution scanning electron microscopy (HR-SEM) images provide detailed structure information for the GSE and FEM analyses. A strongly localized plasmonic mode in the near-infrared spectral region tunable by choice of geometry and with displacement pattern reminiscent of a dark quadrupole mode is identified in the immediate vicinity of the Au subcolumns. We demonstrate tuning of this quadrupole-like mode frequency within the near-infrared spectral range by varying the length of the nano-bamboo subcolumns while maintaining the ratio of Si-Au sub-column-lengths ratio. In addition, a coupled-plasmon-like and Si-descending inter-band transition-like modes are assigned in the visible and ultra-violet spectral regions. Finally, we demonstrate the potential use of the nano-bamboo structures as highly porous plasmonic sensor with optical read out sensitivity at few parts-per-million solvent levels in water. 

\section*{Results and Discussion}

\subsection*{Structural properties}

\begin{figure*}[hbt]
\centering
\includegraphics[width=.9\textwidth]{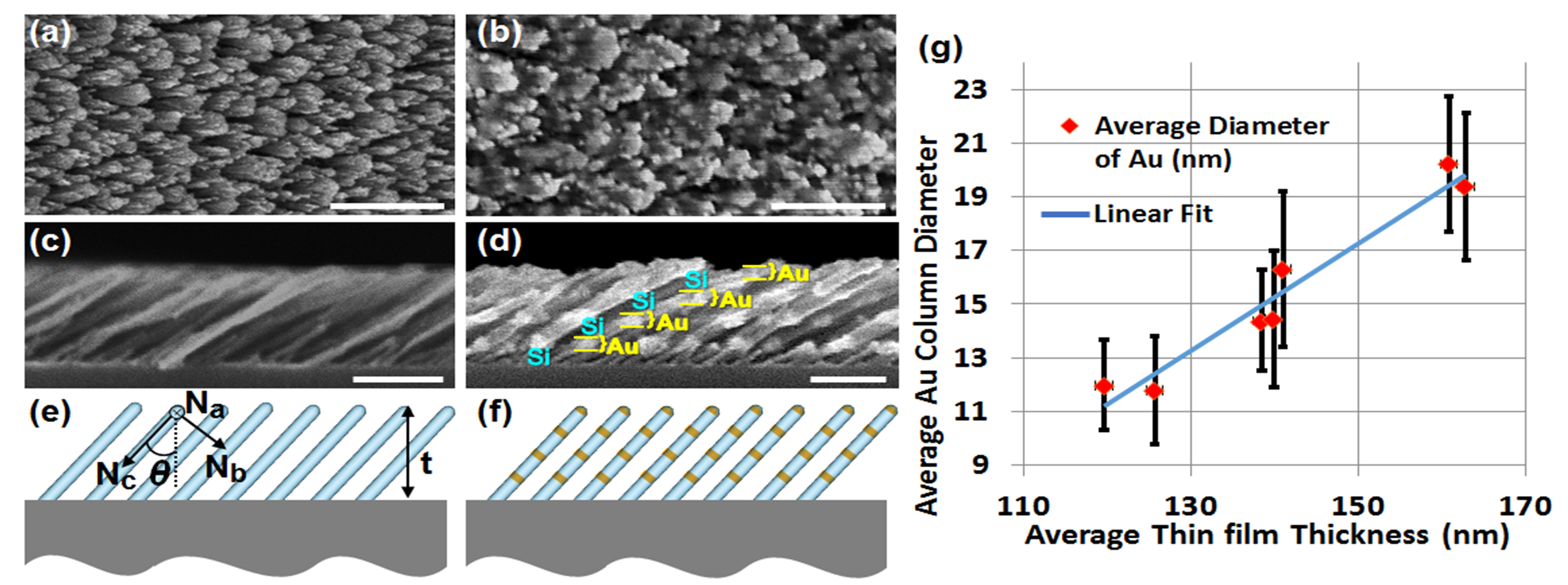}
\caption{Top view (a,b) and cross section (c,d) HR-SEM images, and schematic cross sections (e,f) of a Si SCHTF (a,c,e) with total thickness of 101~nm, and a Si-Au SCHTF (nano-bamboo: (b,d,e)) with total thickness of 148~nm and Si-Au sub-column ratio of approximately 12.8. Scale bars are 100~nm. Overlaid in (e) are the intrinsic orthogonal dielectric polarizability axes, \textbf{N$_a$}, \textbf{N$_b$}, and \textbf{N$_c$}. (g) Evolution of Au sub-column diameter with total thickness of SCHTFs with different Au sub-column lengths obtained from statistical analysis of HR-SEM cross section images. Vertical bars indicate the 90$\%$-confidence standard deviation intervals.}
\label{SEM}
\end{figure*}

Figure~\ref{SEM} shows representative HR-SEM images of our samples, including a Si-SCTF without Au subcolumns. Total thickness of Au and Si sub-column lengths were adjusted by increasing deposition times and by variation of deposition rates for the Au and Si sub-columns. Across all samples investigated here a constant ratio between Si and Au sub-columns of 12.8 was maintained. All samples possessed the same slanting angle, common to all sub-columns, of $\theta$=59$^{\circ}$. We observe that for larger sub-column length, thus for longer deposition times, the column diameters increase slightly. A quantitative analysis of the column widening effect in our samples is depicted in Fig.~\ref{fabricationSCHTF}(g), where HR-SEM image statistic analyses of column diameter versus total Si-Au SCHTF thickness is shown for all samples investigated here. The vertical error bars indicate the standard deviation of the average column diameters found in the different images.

\subsection*{Optical properties}

\subsubsection*{Ellipsometry result: Single Si-Au SCHTF}

\begin{figure}[hbt]
\centering		
\includegraphics[width=0.6\textwidth]{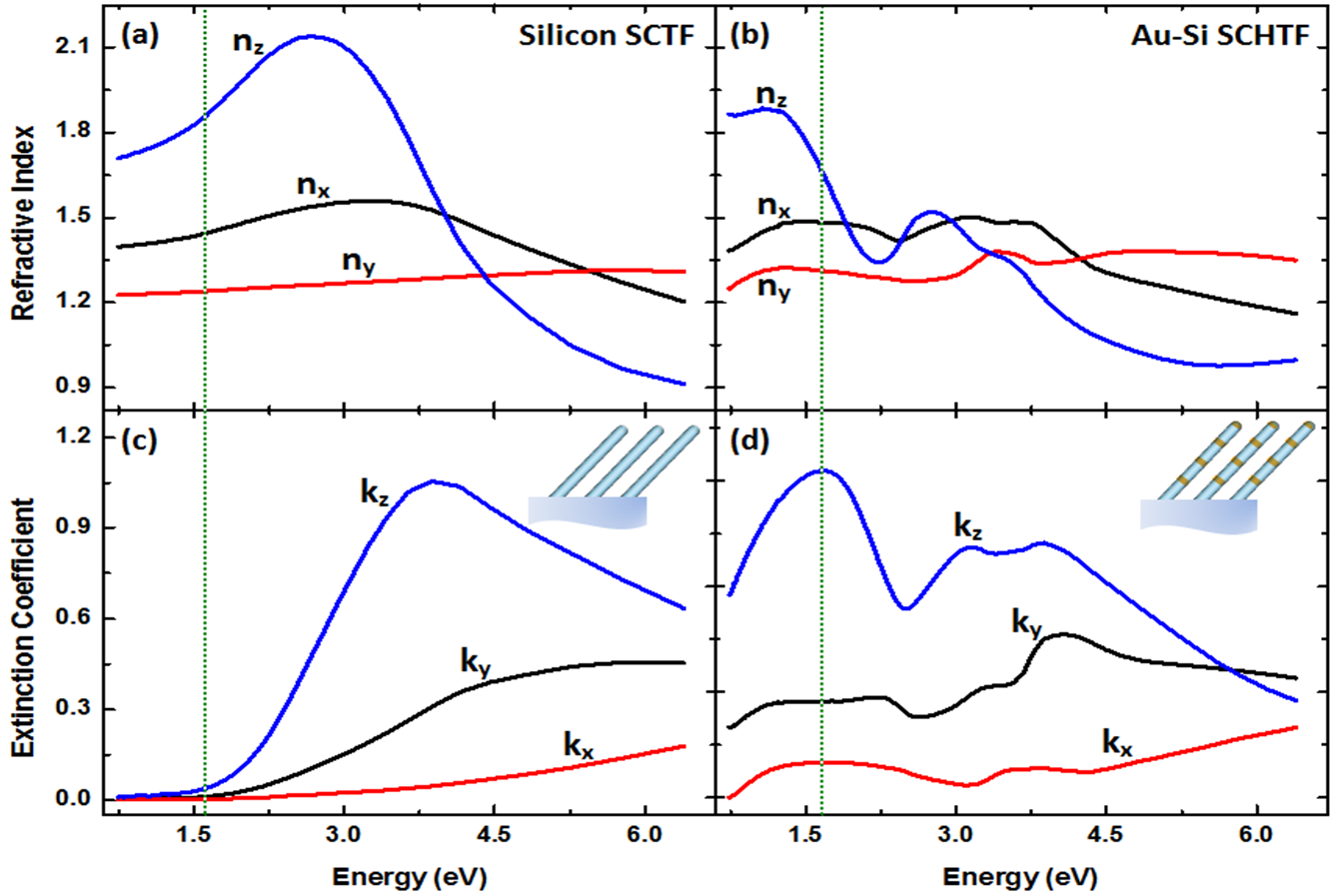}
\caption{Major optical constants of (a,c) Si SCTF and (b,d) Au-Si SCHTF with total thickness of 101~nm and 148~nm, respectively, obtained using the HBLA method and GSE data analysis.}
\label{EXP_GSE_nk}
\end{figure}

Figures~\ref{EXP_GSE_nk}(a-d) show the 3 major optical constants of a Si-Au nano-bamboo structure thin film with total thickness of 148~nm in comparison with those obtained from the pure Si-SCHTF with a total thickness of 101~nm, both determined from GSE analysis using the HBLA formalism (See Methods section). The optical constants correspond to the index of refraction and extinction coefficients for the 3 orthogonal major polarizability axes, $\mathbf{N_a}$, $\mathbf{N_b}$, $\mathbf{N_c}$ (Fig.~\ref{SEM}(e)). For the Si-SCTF, we observe spectra similar to those reported previously.\cite{Schmidt_2013} Index of refraction, $n_c$, and extinction coefficients spectra, $k_c$, for polarization direction parallel to the column axis, $\mathbf{N_c}$, are reminiscent of bulk Si,\cite{Schmidt_2013} and are similar to polycrystalline Si.\cite{JellisonAPLaSi1993} A strong peak is seen in the Si-SCHTF $k_c$ spectrum, which is caused by the strong interband-transition absorption in crystalline silicon attributed to Van Hove singularities in the valance-conduction band joint density of states.\cite{yu1996fundamentals} Without  Au subcolumns, for photon energies below 1.9~eV, the Si SCTF is highly transparent and strongly birefringent, which was noted before.\cite{Schmidt_2013} The optical constants perpendicular to the column axis are smaller due to the intercolumnar spacing and the effective mixture of polarizabilities of void (air) and silicon, which was discussed previously.\cite{Schmidt_2013} In general, it is observed that the intercolumnar spacing within the slanting plane is smaller than perpendicular to the slanting plane, hence, the optical constants for directions $\mathbf{N_b}$ are smaller than for direction $\mathbf{N_b}$.\cite{Schmidt_2013} 

A strong, new absorption peak appears within the extinction coefficient for polarization direction parallel to the column axis at near-infrared photon energies upon the introduction of the Au columnar segments. The spectral position of this new peak is indicated with a vertical line in Figures~\ref{EXP_GSE_nk}(a-d). This peak is completely absent in the Si-SCHTF. The Si inter-band transition-like absorption peak is still present but diminished in amplitude and slightly red shifted. An additional absorption peak occurs in between the new near-infrared peak and the inter-band transition-like peak. Small features are also occuring in the optical constants for polarizarions perpendicular to the column axis, which will be discussed at the end of this paper. 

\subsubsection*{FEM result: Single Si-Au SCHTF}

\begin{table}[ht]
\centering
\caption{Numerical parameters used for FEM calculations in this work.} 
\resizebox{0.5\textwidth}{!}{
\begin{tabular}{ccccc}
\hline\hline
{Parameter} & Value \\ 
\hline\hline
Light incidence angle & $\Phi_{\mathrm{a}}=60^{\circ}$\\
Column slanting angle & $\theta =60^{\circ}$\\
Au subcolumn length ($L_{\mathrm{Au}}$) & $L_{\mathrm{Au}} = (1.75 \dots 5.5)$~nm\\
Si subcolumn length ($L_{\mathrm{Si}}$) & $L_{\mathrm{Si}}=12.75 \times L_{\mathrm{Au}}$\\
Si-Au column radius ($R$) & $R=1.25 \times L_{\mathrm{Au}}+2.5$~nm\\
\hline \hline
\end{tabular}
\label{parameters}
}
\end{table}

\begin{figure}[ht]
\centering
\includegraphics[width=.9\textwidth]{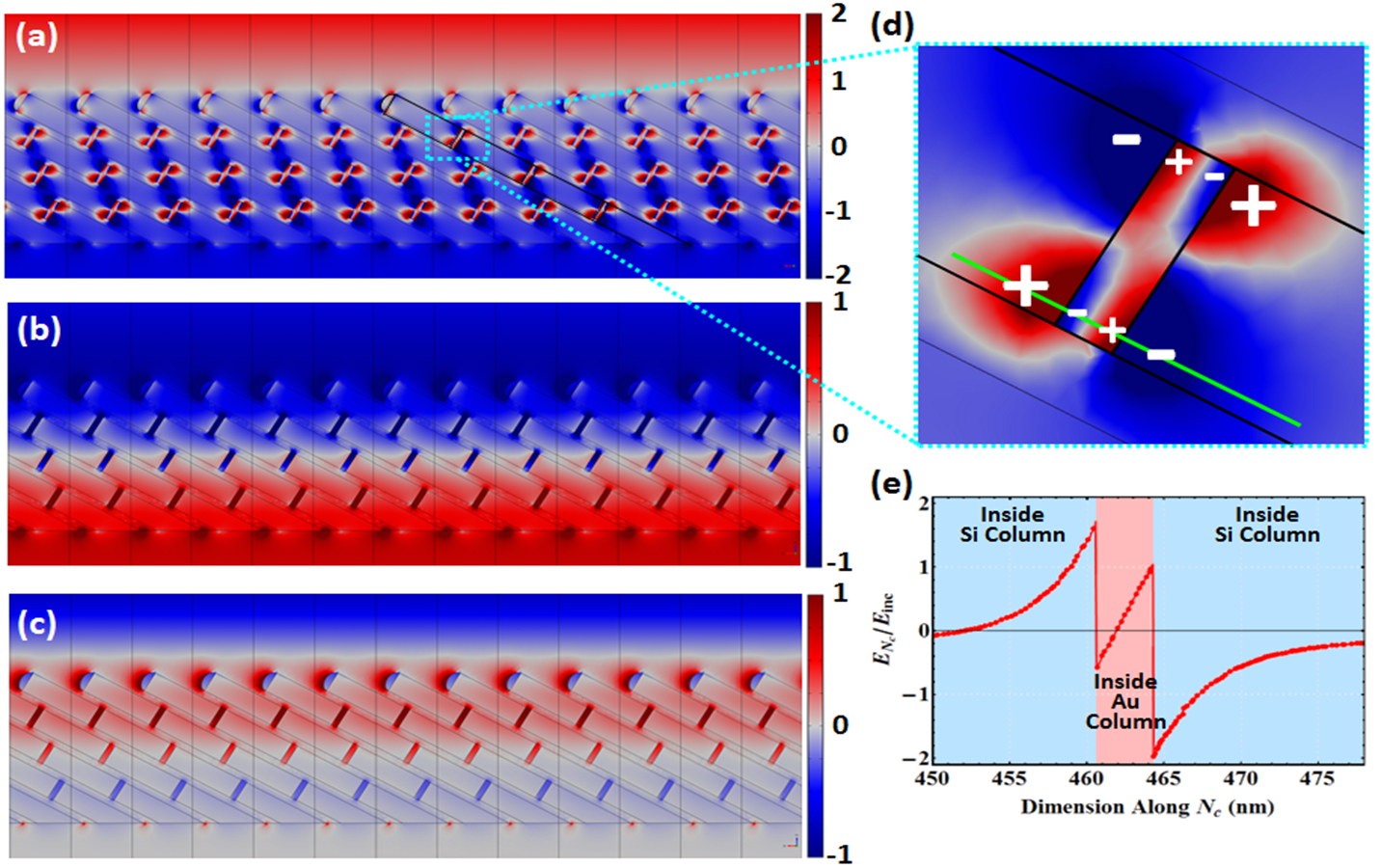}
\caption{(a)-(d): Normalized field enhancement plots of the electric field component parallel to the column axis, $\mathbf{E_{N_c}}$, within the slanting plane of a Si-Au SCHTF calculated using finite element modeling. Parameters are given in Tab.~\ref{parameters}, and here $L_{\mathrm{Au}}=4$~nm. The incident electric field is parallel to \textbf{N$_c$}, for photon energy (a) 1.45~eV (quadrupole-like mode), (b) 2.56~eV (coupled-plasmon-like mode), and (c) 3.6~eV (Si inter-band transition-like mode). (d) Enlarged region near a Si-Au-Si subcolumn sequence for the quadrupole-like mode,  where resemblance with a quadrupole field distribution can be seen. (e) Electric field ratio between incident field amplitude and field amplitude within the columnar structure along the solid green line in (d).}
\label{2x_FEM__size_study}
\end{figure}

Figures~\ref{2x_FEM__size_study}(a-e) depict normalized color density plots of the electric field enhancement component parallel to the column axis, $\mathbf{E_{N_c}}$, within the slanting plane of a Si-Au SCHTF calculated using finite element modeling.
Table~\ref{parameters} lists all numerical parameters used for the FEM calculations. Structural parameters are matched to SEM results. The electric field is shown for photon energies (a) 1.45~eV, (b) 2.56~eV, and (c) 3.6~eV. These  energies match approximately the spectral positions of the experimentally observed peaks in the extinction coefficient for polarization parallel to the Si-Au SCHTF column axis shown in Fig.~\ref{EXP_GSE_nk}(b,d). The field distributions differ drastically, and are used here to identify the origin of the observed peaks in our GSE data analysis. Figure~\ref{2x_FEM__size_study}(a) reveals a highly unusual mode of strongly varying electric field across the spatially-coherent columnar nanoplasmonic superlattice-type thin films. The electric field appears folded like bow-ties within the columnar superlattice-type structures. Upon closer inspection of Fig.~\ref{2x_FEM__size_study}(d), the field distribution projected onto the slanting angle plane closely resembles the near-field pattern of a quadrupole source,\cite{ginzburg2012non, chern2007particle, liu2015tunable} hence, we refer here to this mode as a quadrupole-like mode. The mode at 2.56~eV (Fig.~\ref{2x_FEM__size_study}(b)) resembles a plasmonic mode between top and bottom of the nanoplasmonic superlattice-type thin film. While the top Au subcolumns show a normalized unit-less E-field magnitude along \textbf{N$_c$} axis of -1 (dark blue color), the bottom Au subcolumns reveal a normalized electric-field magnitude of +1 (dark red). Therefore, a net dipole moment is obtained along \textbf{N$_c$} axis. The field distribution is reminiscent of a charge transfer mode with transfer parallel to the column axis. However, the Si subcolumns are treated as insulating in our calculations, and we expect the Si subcolumns in our Si-Au SCHTF samples to be low electrically conductive, as well. Hence, the charge transfer is mediated by an effective Maxwell displacement current across the Si subcolumns. We thus refer to this mode as a coupled-plasmon-like mode. The mode at 3.6~eV (Fig.~\ref{2x_FEM__size_study}(c)) localizes the electric field mostly within the Au subcolumns, and we refer to this mode as the Si inter-band transition-like mode due to its spectral proximity to the Si bulk inter-band transition.\cite{chern2007particle}

\subsubsection*{Ellipsometry result: Si-Au SCHTF thickness evolution}

\begin{figure}[hbt]
\centering		\includegraphics[width=0.95\textwidth]{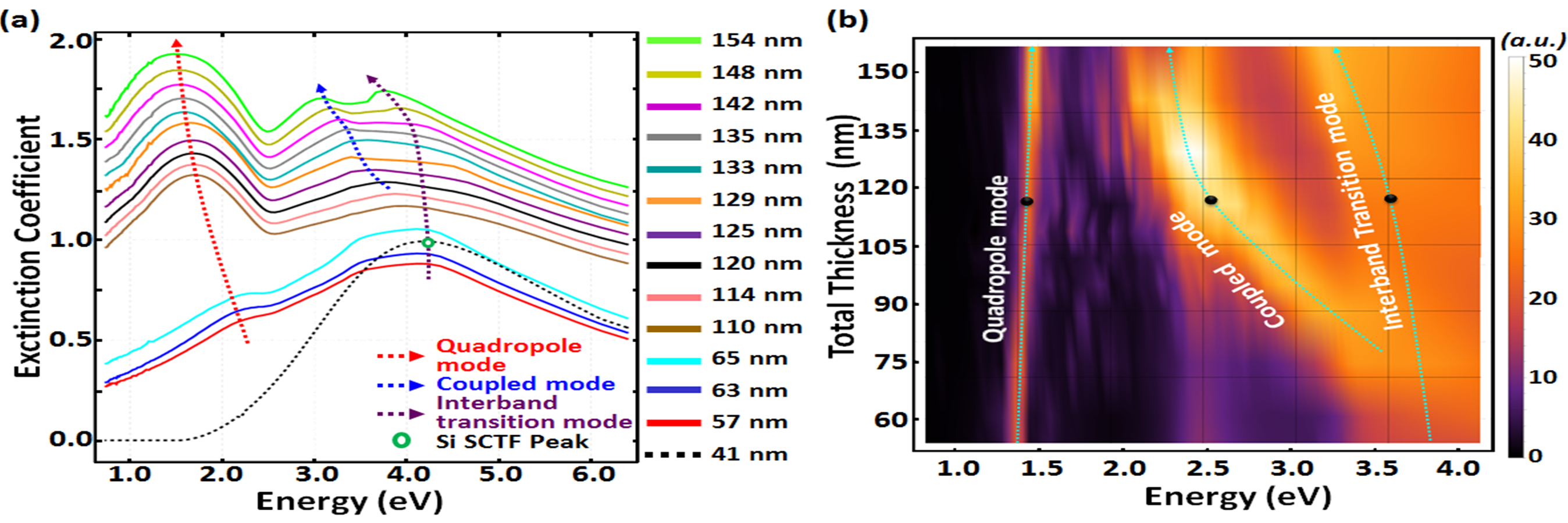}
\caption{ (a) Spectral dependencies of the extinction coefficient for polarization parallel to the column axis (\textbf{N$_c$}) as a function of total Si-Au SCHTF thickness determined from GSE and HR-SEM analyses. Lines are incrementally shifted vertically by 0.05 for convenience. The black dotted line is for a pure Si SCTF with total thickness of 41~nm. Red, blue, and purple dashed arrows indicate the evolution of the quadrupole-like mode, the coupled-plasmon-like mode, and the inter-band transition-like mode, respectively. (b) Surface color rendering of the FEM computed extinction coefficient for polarization parallel to the column axis (\textbf{N$_c$}) as a function of total Si-Au SCHTF thickness. Cyan dashed arrows indicate the evolution of characteristic absorption peaks with total thickness.}
\label{EXP_FEM_extinction}
\end{figure}

Figure~\ref{EXP_FEM_extinction}(a) depicts the evolution of the extinction coefficient for polarization parallel to the column axis, \textbf{N$_c$}, determined by GSE for our samples as a function of total Si-Au SCHTF thickness value. Note that the increase in total thickness results in an increase in total Au content, because we maintain a constant Si-Au subcolumn-length ratio of approximately 12.8 during GLAD deposition. We observe the formation of the Au-subcolumn induced quadrupole-like mode in the spectral range of $\approx$1.4-1.5~eV in all samples. As seen in Fig.~\ref{EXP_FEM_extinction}(a), indicated by a red dashed line, the strength of the peak increases with total thickness. The spectral position of this peak red shifts with increasing thickness at small total thickness values and then appears to become thickness independent for larger thickness values. The Si inter-band transition peak, indicated by a green hollow circle in Fig.~\ref{EXP_FEM_extinction}(a) for a pure Si SCTF, can be recognized in all Si-Au SCHTFs. For small thickness, i.e., for small Au content, the peak remains at the Si inter-band transition energy, and reveals a red shift for larger Au content (larger total thickness). The coupled-plasmon-like mode can be identified by the peak with spectral position intermediate to the quadrupole-like mode and the inter-band transition-like mode in the spectral range of $\approx$3.0-3.5~eV, and which is indicated by the blue dashed line in Fig.~\ref{EXP_FEM_extinction}(a).

\subsubsection*{FEM result: Si-Au SCHTF thickness evolution}

Figure~\ref{EXP_FEM_extinction}(b) depicts the evolution of the extinction coefficient for polarization parallel to the column axis, \textbf{N$_c$}, determined by FEM computations as a function of total Si-Au SCHTF thickness. Table~\ref{parameters} lists all numerical parameters used for the calculations. The experimentally observed effect of slight column ``fanning'' with increasing GLAD deposition time is considered here by introducing a linear increase in column diameter with increasing column length. The Au subcolumn length, $L_{\mathrm{Au}}$, is used as parameter to scale the Si subcolumn length, $L_{\mathrm{Si}}$, and column radius, $R$. A constant ratio between $L_{\mathrm{Si}}$ and $L_{\mathrm{Au}}$ of 12.75 is set for the calculations in order to closely replicate the geometry of our samples.  

The surface color rendering reveals a spectrally narrow, stong absorption peak in the near-infrared region. This mode is the above identified quadrupole-like mode. In Fig.~\ref{EXP_FEM_extinction}(b), the quadrupole-like mode reveals a small blue shift with increasing total thickness. For small thickness (small total Au content), the interband-transition-like peak is close to the Si transition energy, and forms a broad band whose edge is red-shifted for larger thickness. For intermediate thickness values, the coupled-plasmon-mode emerges from the inter-band transition-like mode. Lines are superimposed onto Fig.~\ref{EXP_FEM_extinction}(b) for the three modes discussed in order to guide the eye.

\subsubsection*{Comparison GSE vs FEM result: Si-Au SCHTF thickness evolution}

\begin{figure}[hbt]
\centering		\includegraphics[width=.9\textwidth]{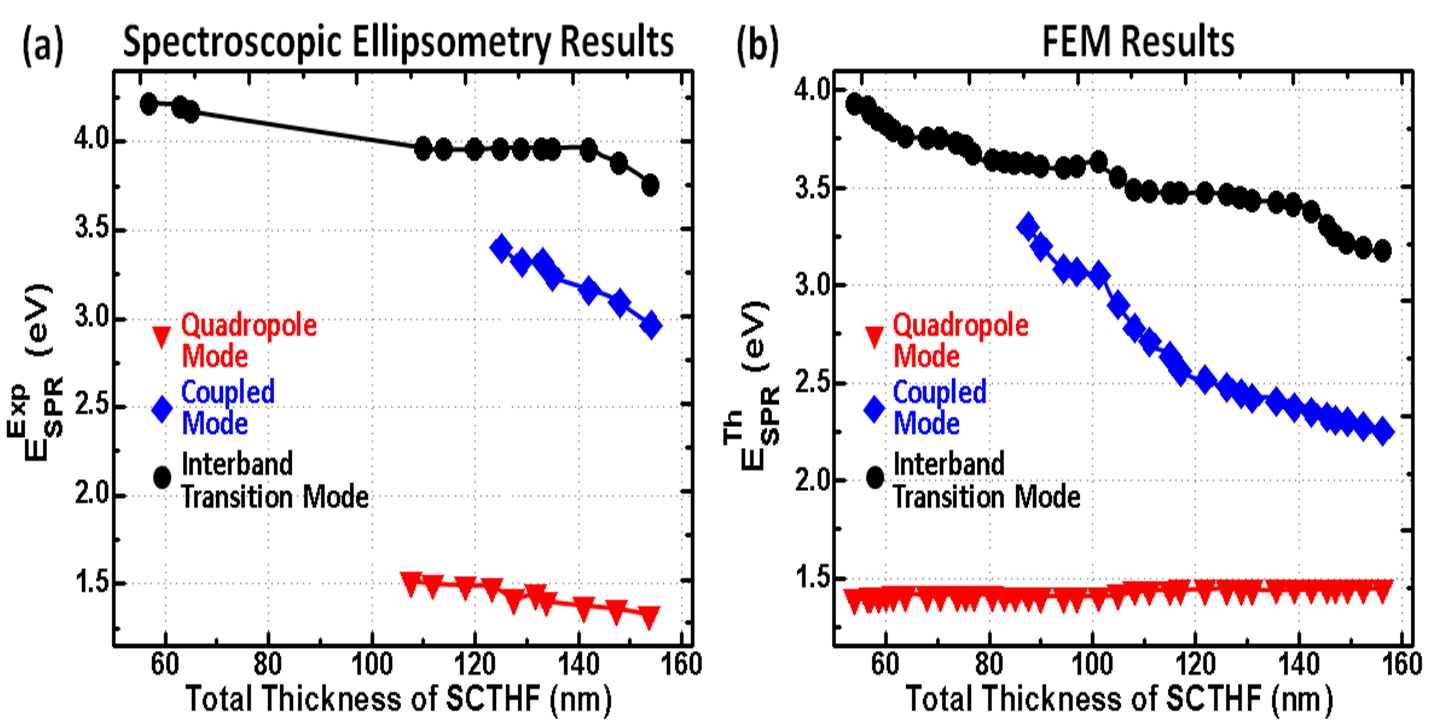}
\caption{Spectral positions of modes identified by GSE analysis (a) and by FEM analysis (b) from the spectral peak behavior of the extinction coefficient for electric field direction parallel to the Si-Au SCHTF column axis, as a function of total Si-Au SCHTF thickness. The ratio between Si and Au subcolumn length in all samples is constant and approximately 12.8.}
\label{peaks}
\end{figure}

Figures~\ref{peaks}(a), and (b) show the spectral positions of the three modes identified in Figure~\ref{EXP_FEM_extinction}(a)  and (b) by ellipsometry analysis, and by computational modeling analysis, respectively, from the spectral peak behavior of the extinction coefficient for electric field direction parallel to the column axis, as a function of total Si-Au SCHTF thickness. An excellent agreement can be seen between the results from our experimental and computational model approaches. The quadrupole-like mode reveals very little, if any, dispersion with increasing thickness in the computational result while a small red-shift is seen from the experiment. We assign this slight disagreement to small structural disorder in the real SCHTF structures versus the assumed perfect geometry in the FEM calculations. During our computational efforts we noted that the quadrupole-like mode is very sensitive to the ratio between the Si and Au subcolumn lengths, hence, to the Si-Au volume ratio. It is for this reason that the ratio was kept constant during the GLAD deposition process. Small variations in Si-Au volume ratio cause subtle shifts of the quadrupole-like mode to lower or higher photon energies, and which thereby can be used as parameter to tune the quadrupole-like mode frequency. The absolute values of the mode energies are slightly different between ellipsometry and FEM results, for example, the inter-band transition-like mode is red shifted overall in the FWM result. This discrepancy may be due to the fact that the spectral dependence of the dielectric constants measured from non-porous polycrystalline Si thin films were used as FEM input parameter for the Si subcolumns, while their true dielectric functions may be slightly different.

\subsubsection*{FEM result: Si-Au subcolumn parameter evolution}

\begin{figure}[hbt]
\centering		\includegraphics[width=.9\textwidth]{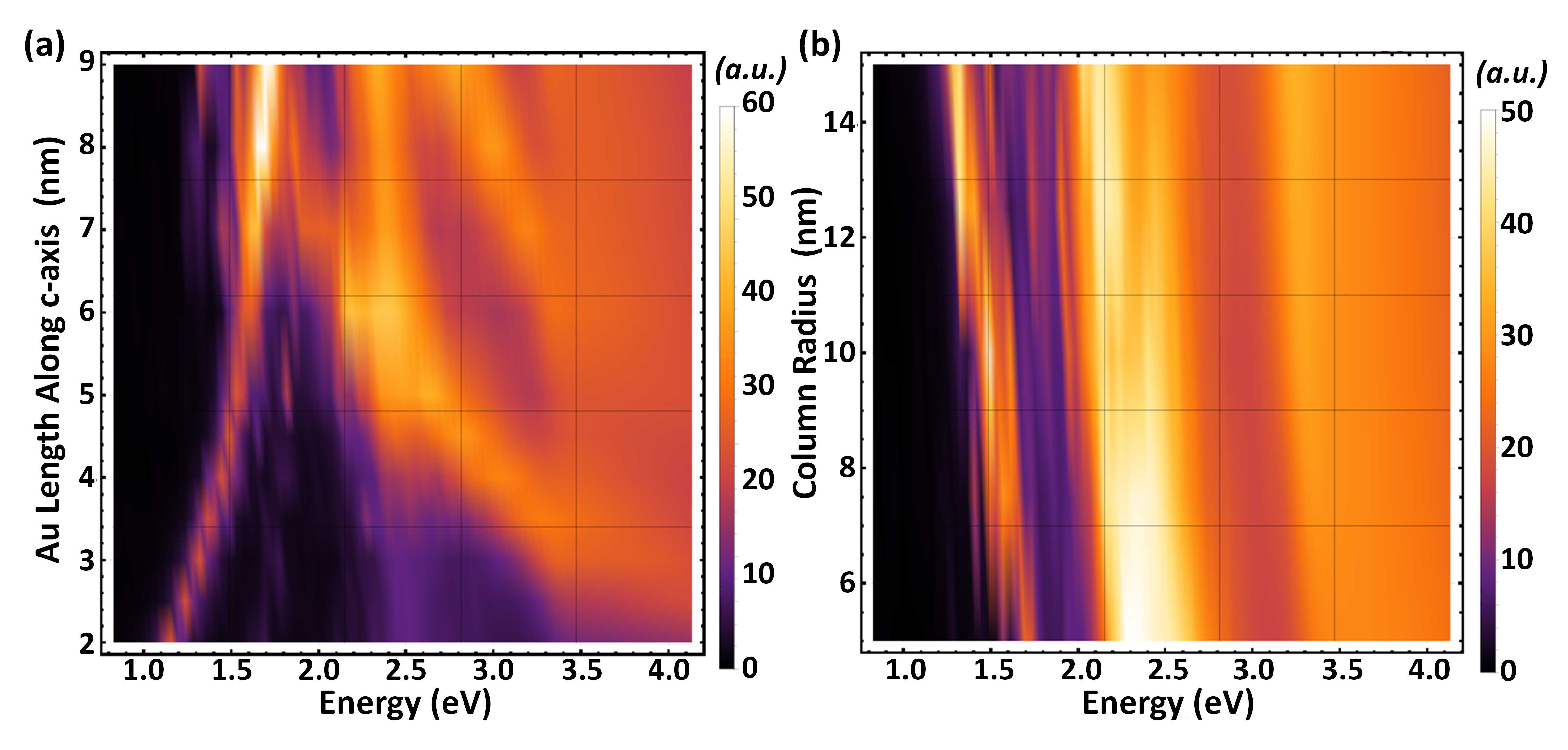}
\caption{(a) Surface color rendering of FEM computed extinction coefficient for polarization parallel to the column axis (\textbf{N$_c$}) as a function of increasing Au subcolumn length, $L_{\mathrm{Au}}$, at constant column radius, $R=7.5$~nm, and constant Si subcolumn length, $L_{\mathrm{Si}}=60$~nm, and (b) for increasing column radius, $R$, at constant Si, $L_{\mathrm{Si}}=60$~nm, and Au, $L_{\mathrm{Au}}=5$~nm, subcolumn length parameters.}
\label{fig:volumestudy}
\end{figure}

Figure~\ref{fig:volumestudy} depicts FEM computed extinction coefficient for polarization parallel to the column axis (\textbf{N$_c$}) as a function of Si-Au subcolumn parameter evolution. Here, in (a) the volume ratio is changed by increasing the Au subcolumn length parameter while keeping all other parameters constant. In (b), the volume ratio is kept constant while the column radius is increased. As it can be seen, all modes observed in our experiments strongly depend on the Si-Au subcolumn parameters. A decrease in Si-Au volume ratio causes a strong blue shift of the quadrupolar mode. The plasmon-coupled like mode appears only for sufficient Au content and strongly couples with the Si inter-band transition like mode (Fig.~\ref{fig:volumestudy}(a)). A second interband transition-like mode appears for larger Au subcolumn length. The increase in column diameter causes a steady red shift of all modes. The combination of the two parameter effects is observed in our experiment, where with increasing total SCHTF thickness the diameter of the subcolumns slightly increases with deposition time, and therefore, we observe only a very small photon energy shift of the quadrupole-like mode in our experiment because both blue shift due to Au content increase and red shift due to column diameter increase compensate. Nonetheless, Figs.~\ref{fig:volumestudy} (a,b) demonstrate the large photon energy tunability of the plasmonic modes with the nano-bamboo SCHTF structures. 

\subsection*{Discussion}

It was proposed previously to exploit plasmonic modes as means to effectively trap and transport incident light by guiding optical radiation along chains of plasmonic centers with suppressed far-field losses. \cite{liu2015tunable} By exploiting such ``dark'' modes as waveguides, efficient light transporting means could be established on the nanoscale.\cite{liu2015tunable} In recent reports\cite{ginzburg2012non, chern2007particle,liu2015tunable} such modes were observed at $\approx$3.5 eV. The quadrupole-like mode observed here appears at much longer wavelength despite the very small Au subcolumn dimensions. The universal plasmon length in Eq.~\ref{eq:universalplasmonlength} predicts the plasmon energy of the Au nanoparticles in our Si-Au SCHTFs to be at $\approx 2.3$~eV, however, we observe a strong absorption at $\approx 1.5$~eV. Note that according to Eq.~\ref{eq:universalplasmonlength}, our Au subcolumn length should be $L_p \approx 280$~nm. The observed plasmonic resonances are in the near infrared due to the strong electromagnetic field coupling within the spatially-coherent sub-wavelength dimension columnar nanoplasmonic superlattice-type thin films. Observations of quadrupolar-like modes were discussed in previous studies. Generally, it was seen that a dipolar mode occurs at lower photon energies  and a quadrupole mode associated with the plasmonic nanoparticle arrangement occurs at higher photon energy.\cite{rodriguez2006seeded, sakai2016quadrupole, yong2014ultrahigh, yi2016dipole} We observe here that a quadrupolar-like mode occurs at lower energy values (1.4~eV-1.5~eV) and that the associated dipolar modes form at higher energy values at $\approx$2.5-3.0eV (coupled-plasmon-like mode) and $\approx$3.5-4.0eV (inter-band transition-like mode)). We assign the observations of the modes in our structures to plasmon-supported interference in thin films with 3-dimensional sub-wavelength ordered geometries. It is thus clear that other interference modes with polarization direction perpendicular to the columnar axis must exist. While we have not further investigated their spectral locations and appearances here, additional modes can be seen as absorption peaks in the major extinction coefficients for polarization perpendicular ($k_x$, $k_y$) in Fig.~\ref{EXP_GSE_nk}(d). We believe that the concept of preparing plasmonic SCHTF bears large potential for designing and exploiting electromagnetic local field enhanced resonances whose frequencies and polarization properties may be varied over large wavelength regions. 

\subsection*{Sensing Refractive Index Change} 

\begin{figure}[hbt]
\centering
\includegraphics[width=0.65\textwidth]{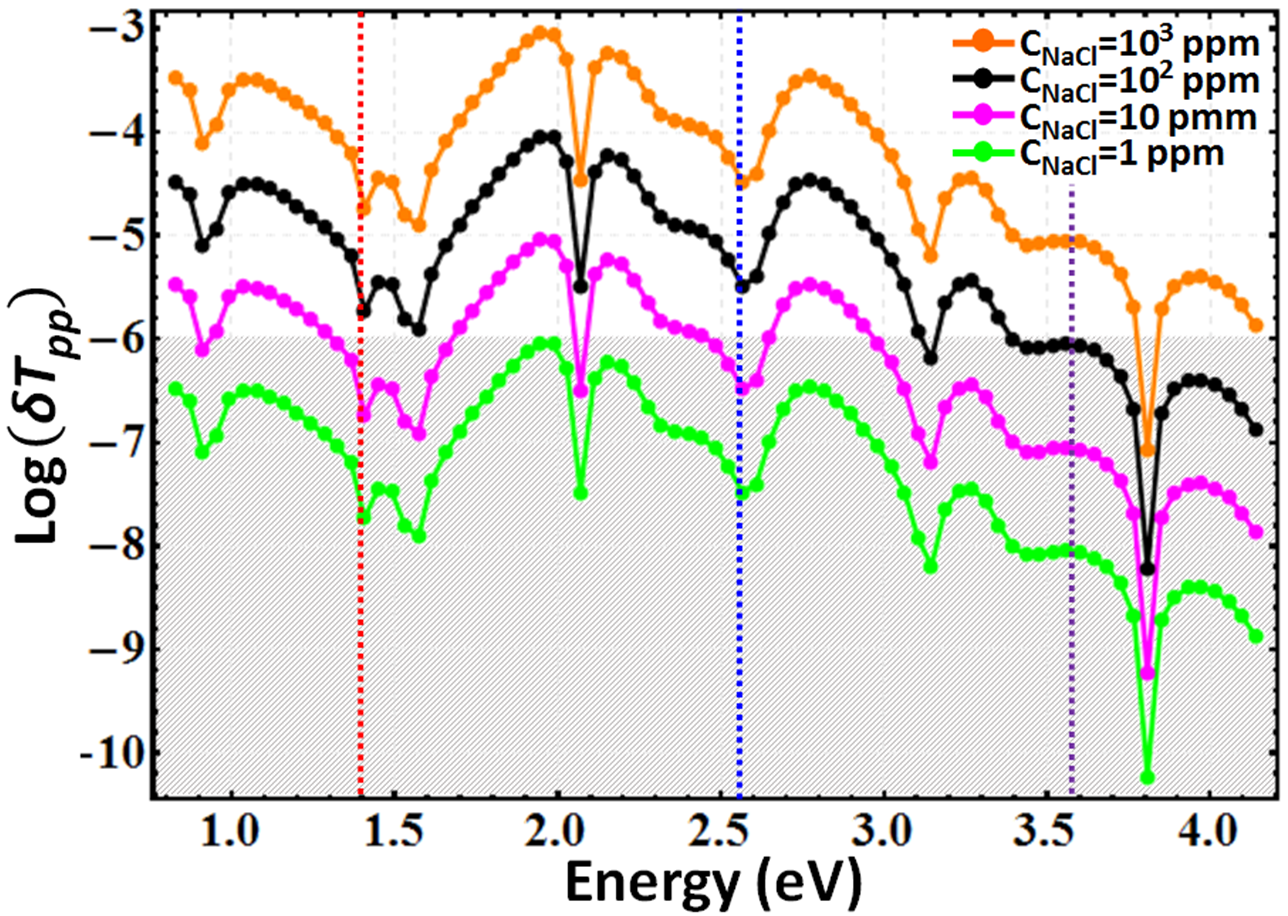}
\caption{Variations of the $p$-polarized incident light to $p$-polarized transmitted light intensity coefficient $T_{ \mathrm{pp }}$ of the nano-bamboo structure immersed in solution upon changes of sodium concentration in nanopure water. Data are obtained from FEM computations, and the Au subcolumn length is $L_{\mathrm{Au}}=4$~nm. The angle of incidence is 30$^{\circ}$, and the slanting plane of the nano-bamboo structure is parallel to the plane of incidence. The gray shaded area corresponds to the usual detection limit of 1$\times$10$^{-6}$ in intensity measurement instrumentation, and is approximated here as the limit at which changes in intensity may be detected. The  vertical lines indicate the photon energies of the SCHTF modes.} 
\label{refractiveindexchange}
\end{figure}

A potential technological application of plasmonic nano-bamboo structures for gaseous or liquid sensing devices is due to the possible combination of ultra-high sensitivity together with device miniaturization.\cite{navas2015laser,willets2007localized} For example, in a recent study, a Si split-ring metasurface was used to sense refractive index change from 1 to 1.2 (equivalent to 20$\%$=2$\times$10$^{5}$ ppm) by observing the shift in the transmission coefficient peaks.\cite{ liu2018high} In another relevant study, a sequential laser ablation method was employed in order to fabricate bimetallic Ag-Au and Ag-Cu core-shell nanoparticles, which were used to obtain different responses to the surrounding refractive index variations, for example, from normal air ambient to full submersion in water.\cite{navas2015laser} A similar study demonstrated sensing refractive index changes by monitoring the response of different shapes of Au nano-rods to their refractive index environment. Measurements followed changes from 1.33 to 1.37, equivalent to the addition (solution) of small inorganic molecules to nanopure water up to $\approx$3$\%$=3$\times$10$^{4}$ ppm.\cite{yong2014ultrahigh} 

Here, we demonstrate performance of the nano-bamboo structures as liquid index of refraction sensor. We monitor the polarized spectral transmission behavior while systematically changing the dielectric properties of the surrounding environment. We investigate ultra-high sensitivity while following the variation of the index of nanopure water upon addition of small inorganic molecules (NaCl) according to levels from 1000 to 1 part-per-million (ppm). We use our FEM model and compute difference transmission spectra for incident linearly polarized light ($p$-polarized), where differences are taken between the transmittance of the nano-bamboo structure immersed in nanopure water with concentrations of added molecules, $c_{\mathrm{NaCl}}$, and the transmittance of the SCHTF while immersed in pure water only:

\begin{equation}
\delta T_{ \mathrm{pp }}   =T_{ \mathrm{pp }} ( c_{ \mathrm{NaCl}})-T_{ \mathrm{pp }} (c_{\mathrm{ NaCl}}=0),
\label{refractiveindexchange1}
\end{equation}

\noindent where $T_{ \mathrm{pp }}$ is the $p$-polarized incident light to $p$-polarized transmitted light intensity coefficient. The angle of incidence is 30$^{\circ}$, and the slanting plane of the nano-bamboo structure is parallel to the plane of incidence. Hence, the incident electric field is parallel to the column axis. The refractive index of the salt-water solution, $n$, is calculated by using the Arago-Biot approach:\cite{reis2010refractive, sharma2007density}

\begin{equation}
n=n_{\mathrm{H_{2}O}}+(n_{\mathrm{NaCl}}-n_{\mathrm{H_{2}O}}) c_{\mathrm{NaCl}},
\label{maxwelleq1}
\end{equation}

\noindent where $n_{\mathrm{H_2O}}=1.33$ is the water refractive index, $n_{\mathrm{NaCl}}=1.57$ is the (dry) salt refractive index, and $c_{\mathrm{NaCl}}$ is the salt concentration in the solution. 

Figure~\ref{refractiveindexchange} depicts the change in the p-polarized transmission coefficients for different C$_{NaCl}$ values, from 1 ppm to 1000 ppm. The gray area indicates the region where a photon-counting based spectrophotometer typically will report noise, with a linear dynamic range over 6 orders of magnitude.\cite{peev2016anisotropic} Under such assumptions, it can be seen that the existence of 1 ppm NaCl in pure water will be detectable. While this sensitivity is very high, the specificity to a certain solvent maybe low. However, absorbing organic molecules may alter the interference features within the transmittance spectrum differently which can be the subject of future work.

\section*{Conclusions}

Fabrication of spatially-coherent columnar plasmonic nanostructure superlattice-type thin films by subsequent and repeated depositions of silicon and gold leads to nanometer-dimension subcolumns with controlled lengths and nanoplasmonic properties. Resembling nano-bamboo-like structures the superlattice-type thin films contain very small gold junctions sandwiched between larger silicon column sections. A combined generalized spectroscopic ellipsometry and finite element method analysis is performed to elucidate the strongly anisotropic optical properties of the highly-porous nano-bamboo structures. We observe and identify strongly localized plasmonic modes with displacement pattern reminiscent of a dark quadrupole mode. Additional coupled-plasmon-like and inter-band transition-like modes occur in the visible and ultra-violet spectral regions, respectively. We elucidate an example for the potential use of the nano-bamboo structures as a highly porous plasmonic sensor with optical read out sensitivity to few parts-per-million solvent levels in water.

\section*{Methods}
\subsection{Glancing Angle Deposition (GLAD)}
Using a custom built ultra-high vacuum GLAD system with a base pressure of 1.0 $\times$ 10$^{-9}$ mbar, periodic alternating layers of Si and Au were deposited at oblique angles onto Si(100) substrates. The p-type Si substrates  were determined to have a native oxide layer of approximately 1.8~nm prior to deposition. An electron beam evaporation was utilized and the particle flux of the material was directed towards the substrate at an angle of 85$^{o}$ from the normal. An electron beam of 8.8~kV and 180~mA was used to evaporate the Si while for the Au 8.9~kV and 215~mA were used. These parameters resulted in an average chamber pressure of 2.0$\times$ 10 $^{-7}$ mBar for Au growth and 1.2 $\times$ 10$^{-7}$~mbar for Si growth. A quartz crystal micro-balance (QCM) was used to measure the rate and total amount of deposited material. Multiple samples were deposited with different total thickness where we aimed to maintain a constant Si-Au sub-column-lengths ratio. A ratio of approximately 12.8 was obtained. Sub-column lengths, and thereby total thickness of the SCHTFs, were controlled by variation of deposition rate and deposition time. Typical deposition rates were 0.2~\AA/s for Au and 2.5~\AA/s for Si. An electronic shutter system was utilized to grow each periodic layer for the same time interval. The Au growth was controlled by the total amount of Au deposited on the QCM for a better control over the thickness of these extremely thin layers. All SCHTF structural parameters (thickness, sub-column lengths, slanting angle) were determined from high-resolution scanning electron microscopy (HR-SEM) images.  

\subsection{Generalized Spectroscopic Ellipsometry (GSE)}
Generalized Spectroscopic Ellipsometry (GSE)\cite{SchubertJOSAA13_1996} can be used to obtain the optical properties of anisotropic materials. In this process, the so-called Mueller matrix\cite{MuellerMIT1943} is obtained, which represents the linear optical properties of the sample under investigation. The Mueller matrix represents a real valued 4$\times$4 matrix which describes the effect of the sample onto an incoming electromagnetic wave represented by its Stokes vector into its outgoing Stokes vector. The off-block diagonal Mueller matrix elements are particularly sensitive to anisotropy. An appropriate physical model must be determined and a best-match model calculation must be employed in order to extract physically meaningful parameters from measured generalized ellipsometry data.\cite{Schubert96,Fujiwara_2007} Details of the numerical procedures were discussed previously.\cite{Schmidt_2013} \textit{Ex-situ} GSE Mueller matrix measurements were conducted in the spectral range of 0.72-6.2~eV using a dual rotating compensator ellipsometer (RC2, J.A.~Woollam~Co.,~Inc.). All spectra were collected at angles of incidence of 45$^\circ$, 55$^\circ$, 65$^\circ$, and 75$^\circ$. At each angle of incidence ($\phi_A$), spectra were measured over a full azimuthal rotation (ie. from 0$^\circ$ to 360$^\circ$ by 6$^\circ$ increments). All ellipsometric spectra were analyzed using the HBLA method and the ellipsometry data model software (WVASE32, J.A.~Woollam~Co.,~Inc.).

\subsection{Homogeneous Biaxial Layer Approach (HBLA)}
Previous analysis reports from single-material SCTFs, core-shell SCTFs\cite{Mock_2016, SchmidtPassivation2012} and SCHTFs\cite{Sekora_2016} introduced the homogeneous biaxial layer approach (HBLA).\cite{Schmidt_2013} The HBLA method assumes that the linear optical response of SC(H)TFs can be described by a single homogeneous layer permitting line-shape function independent wavelength-by-wavelength determination of effective major dielectric functions along three major dielectric polarizability axes, \textbf{N$_a$}, \textbf{N$_b$}, and \textbf{N$_c$} (Fig. \ref{SEM}). In addition to the three major functions, wavelength-independent structural parameters such as slanting angle $\theta$ and SCHTF thickness $t$ are obtained from the best match model calculations.\cite{Schmidt_2013} All samples in this work area optically analyzed by using the HBLA for best-match model calculation of the anisotropic optical properties measured by GSE. As a result we report the anisotropic effective dielectric functions determined by HBLA.  

\subsection{Finite Element Modeling (FEM)}
Finite element modeling (FEM) calculations are performed by using the RF module of COMSOL Multiphysics software. We determine the full wave electrodynamic solutions to Maxwell’s equations. Incoming plane wave radiation with a known polarization interacts with Au-Si SCHTF structures and both reflected and transmitted plane waves acquire different polarizations. Three dimensional visualization of the electric field distribution in the vicinity of the columnar structures are obtained, and the spectral dependence of the Jones matrix elements are calculated.\cite{balanis2016antenna,Fujiwara_2007} The multiple port definition enabled the extraction of the S-parameters, or scattering parameters.\cite{balanis2016antenna} The transmission and reflection coefficients are calculated from the scattering parameters, thus the absorption coefficient is obtained, accordingly. All simulations were performed for the angle of incidence, $\theta$=60$^\circ$. A hexagonal close packed unit cell definition with periodic boundary condition is incorporated to account for the interactions between the neighboring columns. A numerical adaptive-length mesh was used with maximum element sizes of $4.75$~nm, and a  parametrized unit cell for the Au-Si SCHTF structure is created in COMSOL Multiphysics environment. Numerical values for the dielectric constants for the Au and Si sub-columns and their wavelength dependencies are taken from experimental data determined on bulk materials.\cite{palikhandbook}  


\section*{Acknowledgements}
This work was supported in part by the National Science Foundation (NSF) through EPSCoR Research Infrastructure Improvement program award EPS-1004094, the Nebraska Materials Research Science and Engineering Center (MRSEC) award DMR-1420645, and awards CMMI 1337856, EAR 1521428, and DMR 18XXXXXX. This work was partly supported by the German Research Foundation award FE 1532/1-1, by Air Force Research Office award FA9550-18-1-0360, and by American Chemical Society award XXXXX. The authors further acknowledge financial support by the University of Nebraska-Lincoln, the J.~A.~Woollam Co., Inc., and the J.~A.~Woollam Foundation.

\section*{Author contributions statement}
R.~F., M.~H., U.~K., and E.~S. performed GLAD deposition of Si-Au nano-bamboo structures. D.~S. and U.~K. performed HR-SEM image analyses. A.~M., U.~K., R.~F., and M.~H. performed GSE analyses. U.~K., C.~A., and M.~S. performed FEM analyses. U.~K., C.~A., and M.~S. wrote the manuscript. The manuscript was edited and approved by all authors. The project was supervised by E.~S., C.~A., and M.~S.

\section*{Additional information}
To include, in this order: \textbf{Accession codes} (where applicable); \textbf{Competing financial interests} (mandatory statement). 

The corresponding author is responsible for submitting a \href{http://www.nature.com/srep/policies/index.html#competing}{competing financial interests statement} on behalf of all authors of the paper. This statement must be included in the submitted article file.

\end{document}